\documentclass[aps,prl,twocolumn,superscriptaddress,letterpaper]{revtex4-1}
\usepackage{graphicx}
\usepackage{amsmath}
\usepackage{esint}
\usepackage{verbatim}
\usepackage{color}
\usepackage{SIunits}
\usepackage{hyperref}

\newcommand{\Tf}{T_{f}}
\newcommand{\nfm}{n_{f,m}}

\newcommand{\Tm}{T_{m}}
\newcommand{\Tr}{T_{r}}
\newcommand{\GammaAS}{\Gamma_{\text{AS}}}
\newcommand{\GammaS}{\Gamma_{\text{S}}}

\newcommand{\SiO}{\text{Si}\text{O}_2}

\newcommand{\omegap}{\omega_{p}}
\newcommand{\omegad}{\omega_{d}}
\newcommand{\ndrive}{n_{d}}
\newcommand{\Pdrive}{P_{d}}
\newcommand{\nprobe}{n_{p}}
\newcommand{\npulse}{n_{\text{pulse}}}
\newcommand{\Deltard}{\Delta_{r,d}}
\newcommand{\omegapulse}{\omega_{\text{pulse}}}
\newcommand{\gzeroE}{g_{\text{0}}}
\newcommand{\kappaE}{\kappa}
\newcommand{\kappaEe}{\kappa_{e}}
\newcommand{\kappaEi}{\kappa_{i}}
\newcommand{\omegacE}{\omega_{r}}
\newcommand{\gammaOME}{\gamma_{\text{EM}}}
\newcommand{\GOM}{G}
\newcommand{\Qr}{Q_{r}}
\newcommand{\Qri}{Q_{r,i}}
\newcommand{\etam}{\eta}
\newcommand{\Cm}{C_{m}}
\newcommand{\Cs}{C_{s}}
\newcommand{\Cl}{C_{l}}
\newcommand{\Ctot}{C_{\text{tot}}}
\newcommand{\gammai}{\gamma_{i}}
\newcommand{\omegam}{\omega_{m}}
\newcommand{\xzpf}{x_{\text{zpf}}}
\newcommand{\meff}{m_{\text{eff}}}
\newcommand{\Qm}{Q_{m}}
\newcommand{\gammam}{\gamma_{m}}

\newcommand{\nwb}{n_{b,\text{wg}}}
\newcommand{\nrb}{n_{b,r}}

\newcommand{\nadd}{n_{\text{add}}}
\newcommand{\nm}{n_{m}}

\begin{document}

\title{Superconducting cavity-electromechanics on silicon-on-insulator}

\author{Paul B. Dieterle}
\affiliation{Kavli Nanoscience Institute and Thomas J. Watson, Sr., Laboratory of Applied Physics, California Institute of Technology, Pasadena, CA 91125, USA}
\affiliation{Institute for Quantum Information and Matter, California Institute of Technology, Pasadena, CA 91125, USA}
\thanks{These authors contributed equally to this work.}
\author{Mahmoud Kalaee}
\affiliation{Kavli Nanoscience Institute and Thomas J. Watson, Sr., Laboratory of Applied Physics, California Institute of Technology, Pasadena, CA 91125, USA}
\affiliation{Institute for Quantum Information and Matter, California Institute of Technology, Pasadena, CA 91125, USA}
\thanks{These authors contributed equally to this work.}
\author{Johannes M. Fink}
\altaffiliation[Current address: ]{Institute of Science and Technology Austria (IST Austria), 3400 Klosterneuburg, Austria}
\affiliation{Kavli Nanoscience Institute and Thomas J. Watson, Sr., Laboratory of Applied Physics, California Institute of Technology, Pasadena, CA 91125, USA}
\affiliation{Institute for Quantum Information and Matter, California Institute of Technology, Pasadena, CA 91125, USA}
\author{Oskar Painter}
\email{opainter@caltech.edu}
\affiliation{Kavli Nanoscience Institute and Thomas J. Watson, Sr., Laboratory of Applied Physics, California Institute of Technology, Pasadena, CA 91125, USA}
\affiliation{Institute for Quantum Information and Matter, California Institute of Technology, Pasadena, CA 91125, USA}  

\date{\today}
\begin{abstract}
Fabrication processes involving anhydrous hydrofluoric vapor etching are developed to create high-$Q$ aluminum superconducting microwave resonators on free-standing silicon membranes formed from a silicon-on-insulator wafer.  Using this fabrication process, a high-impedance $8.9$~GHz coil resonator is coupled capacitively with large participation ratio to a $9.7$~MHz micromechanical resonator.  Two-tone microwave spectroscopy and radiation pressure back-action are used to characterize the coupled system in a dilution refrigerator down to temperatures of $\Tf = 11$~mK, yielding a measured electromechanical vacuum coupling rate of $\gzeroE/2\pi \approx 24.6$~Hz and a mechanical resonator $Q$-factor of $Q_{m}=1.7\times 10^7$.  Microwave back-action cooling of the mechanical resonator is also studied, with a minimum phonon occupancy of $\nm \approx 16$ phonons being realized at an elevated fridge temperature of $\Tf = 211$~mK.    
\end{abstract}
\maketitle

Recent work in the field of cavity-optomechanics has shown the feasibility of using radiation pressure to cool micromechanical objects close to their quantum ground state~\cite{Teufel2011,Chang2011}, to measure the quantum motion of such objects~\cite{Safavi-Naeini2012a,Weinstein2014}, and to prepare non-classical mechanical states using back-action evading techniques~\cite{Wollman2015,Pirkkalainen2015,Lecocq2015a}.  In a dual role, mechanical objects may be used to create large electromagnetic nonlinearities for slowing~\cite{Weis2010,Chang2011,Safavi-Naeini2011,Teufel2011}, squeezing~\cite{Safavi-Naeini2013b,Purdy2013}, or even shifting the frequency of light~\cite{Hill2012}. These experiments have utilized either optical or microwave photons to induce radiation pressure forces, though recent work has coupled opto- and electro-mechanical systems and realized reversible microwave-to-optical conversion~\cite{Andrews2014}. An outstanding problem in the field is to realize such conversion in a fully integrated, on-chip platform~\cite{Davanco2012b,Pitanti2015}. 

\begin{figure}[htp!]
\begin{center}
\includegraphics[width=\columnwidth]{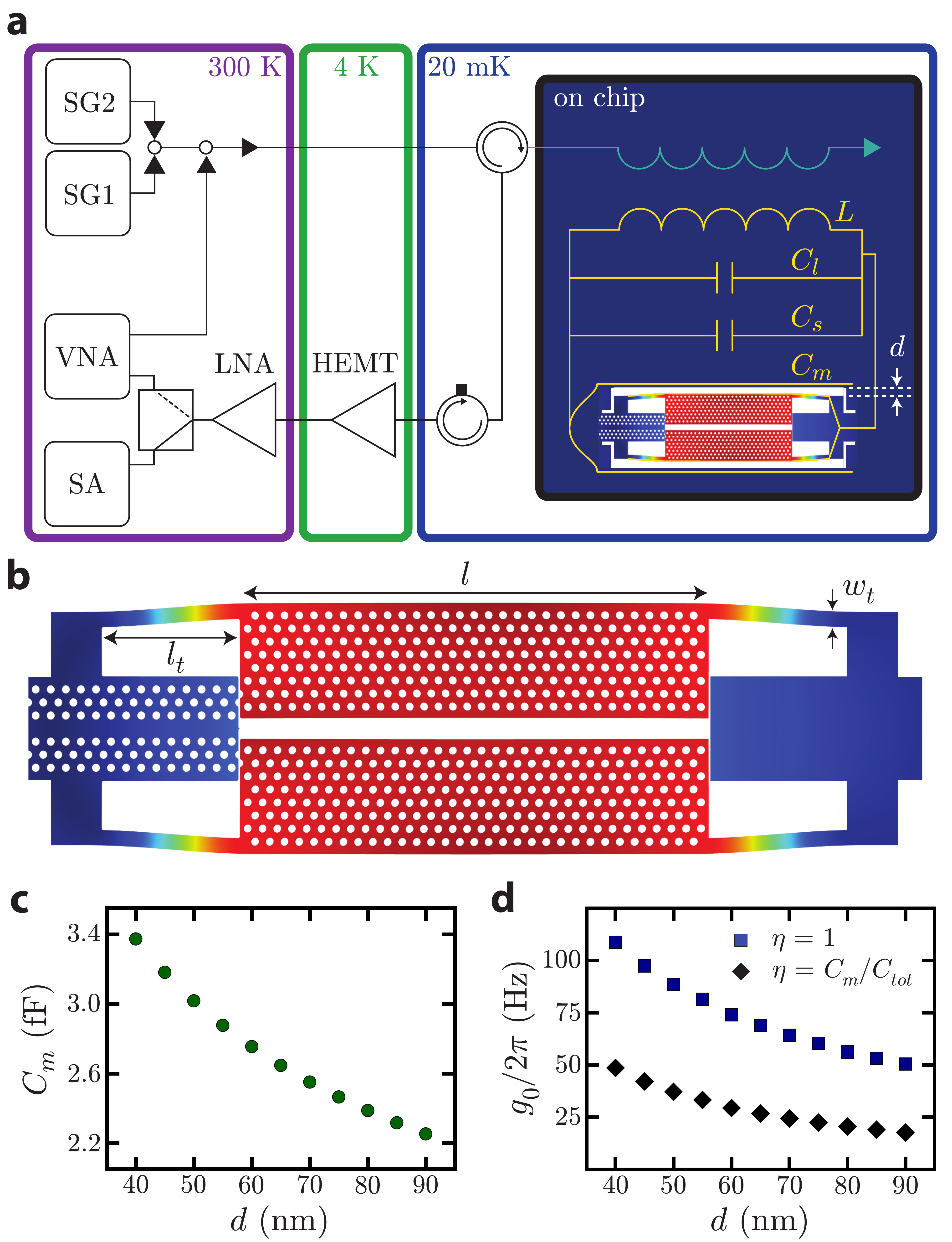}
\caption{\textbf{a}, Schematic of the electromechanical circuit and measurement setup.  The electromechanical circuit (yellow) is inductively coupled to a $2$~$\mu$m wide wire (turquoise) which shorts to ground and reflects the signal. Acronyms: SG$_{i}$ = microwave signal generator, VNA = vector network analyzer, SA = spectrum analyzer, LNA = low-noise amplifier, HEMT = high-electron-mobility transistor amplifier.  \textbf{b}, FEM simulation of the differential mechanical mode. In this work, $l = 13.5$~$\mu$m, $w_t = 440$~nm, and $l_t = 4$~$\mu$m. These values give a simulated mechanical mode frequency of $\omegam/2\pi = 9.76$ MHz.  \textbf{c},  Plot of FEM simulation values of $\Cm$  versus slot size $d$.  \textbf{d},  Plot of FEM simulation values of $\gzeroE$ versus slot size $d$ for: (i) $\Cs = \Cl = 0$~fF corresponding to an ideal $\eta = 1$ (blue squares) and (ii) $\Cl = 3.05$~fF and $\Cs = 1.13$~fF from FEM simulations of the circuit (black diamonds).  For these plots, the resonance frequency is fixed at the measured frequency of $\omegacE/2\pi = 8.872$~GHz.  At the estimated capacitor gap of $d \approx 60$~nm from SEM images, the theoretical values of the motional capacitance and the vacuum coupling rate are $\Cm = 2.76$~fF and $\gzeroE/2\pi = 29.3$~Hz, respectively.} \label{fig:design}
\end{center} 
\end{figure}

Here, we develop a new fabrication process for the creation of high-$Q$ microwave superconducting aluminum (Al) resonators on thin-film silicon membranes suitable for integration with mechanical and photonic devices. As a proof of concept, we demonstrate parametric radiation pressure coupling of an $8.9$~GHz microwave superconducting resonator to the motion of a $9.7$~MHz silicon micromechanical resonator.  The electromechanical circuit, shown schematically in Fig.~\ref{fig:design}(a), consists of a high-impedance microwave coil resonator capacitively coupled to the fundamental in-plane differential mode of a pair of patterned silicon slabs.  Although not a feature exploited in the present study, the patterned slabs also form a slotted photonic crystal cavity which supports an optical resonance in the $1500$~nm telecom wavelength band~\cite{Safavi-Naeini2010,Winger2011,Pitanti2015}.  In principle, this mechanical resonator (what we hereafter refer to as the ``H-slot'' resonator) could simultaneously couple to optical photons in the photonic crystal cavity and microwave photons in the superconducting microwave resonator. 


The H-slot mechanical resonator is depicted in Fig.~\ref{fig:design}(b), where finite-element method (FEM) numerical simulations~\cite{COMSOL} are used to solve for the fundamental in-plane mechanical motion of the structure.  The resonator is formed from a Si layer of $300$~nm thickness, and consists of two patterned slabs that are separated by a central nanoscale slot and tethered on each end to a central clamp point.  As mentioned, the hole patterning in the two slabs produces a localized photonic crystal cavity.  The hole patterning on the left side of the H-slot resonator forms a photonic crystal optical waveguide which can be used to efficiently excite the optical cavity.  Aluminum electrodes are fed into the H-slot resonator from the right side of the structure, and span the outer edges of the two slabs forming one half of a vacuum gap capacitor [labeled $\Cm$ in Fig.~\ref{fig:design}(a)].  The length ($l = 13.5$~$\mu$m) of the photonic crystal slabs is chosen long enough to support a high-$Q$ optical mode and to realize a motional capacitance on the scale of a few femtoFarad.  The width ($w$) of the photonic crystal slabs is chosen to accommodate a number of photonic crystal periods that should (again) provide high optical $Q$, but otherwise is minimized to limit the motional mass of the resonator.  The slab photonic crystals are supported by tethers whose length ($l_{t} = 4$~$\mu$m) and width ($w_{t} = 440$~nm) produce a simulated mechanical frequency of $\omegam/2\pi \approx 9.76$~MHz for the fundamental in-plane differential mode, compatible with resolved-sideband pumping through the coupled microwave circuit.

The simulated effective mass and zero-point amplitude of the differential mode are $\meff = 42.9$~pg and $\xzpf = 4.5$~fm, respectively. These figures include the aluminum wires (width $=250$~nm, thickness $=60$~nm) that form the vacuum gap capacitor.  By using a tuning fork design in which the upper and lower slabs are coupled together at each end through the central tether clamp points, acoustic radiation out the ends of H-slot resonator can be greatly reduced.  Optimization of the tether clamp point geometry yields a numerically simulated mechanical quality factors as high as $\Qm = 3.7\times 10^7$.

\begin{figure*}[htp!]
\begin{center}
\includegraphics[width=\textwidth]{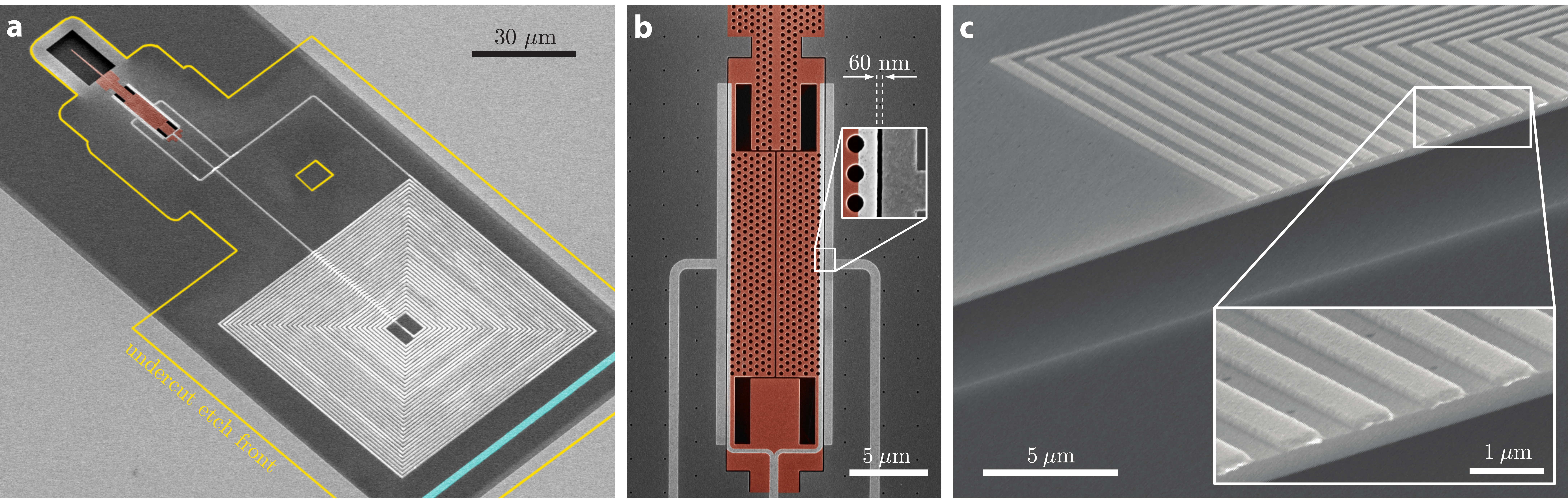}
\caption{\textbf{a}, SEM image of the fabricated microwave coil resonator and H-slot mechanical resonator.  The H-slot resonator region is colored red and the undercut region is outlined in yellow.  The coupling wire is colored turquoise. \textbf{b},  A zoomed in SEM image of the H-slot mechanical resonator.  Inset: a close-up of the $60$~nm wide capacitor gap formed by a $250$~nm wide Al electrode on the photonic crystal slab and a $550$~nm wide Al electrode on the outer Si support membrane. Both wires are $60$~nm thick, as is the ground plane.  \textbf{c}, Cross-section image showing the suspended membrane with a coil on top. The Al forming the coil is $120$~nm thick. The $3$~$\mu$m thick dark area underneath the Si membrane is the undercut region where SiO$_2$ has been etched away. The bottom layer is the Si handle wafer.} \label{fig:fab}
\end{center} 
\end{figure*}

The vacuum electromechanical coupling rate of the H-slot mechanical resonator to the microwave coil resonator is given by,

\begin{equation}
\gzeroE = \xzpf \frac{\partial \omegacE}{\partial u} = - \etam \xzpf \frac{\omegacE}{2\Cm}\frac{\partial \Cm}{\partial u},
\label{eq:gzero}
\end{equation}

\noindent where $u$ is the generalized amplitude coordinate of the fundamental in-plane differential mode of interest, $\xzpf=\sqrt{\hbar/(2\omegam \meff)}$ is the zero-point amplitude of the mechanical resonance, $\etam = \Cm/\Ctot$ is the participation ratio of the motional capacitance ($\Cm$) to the total capacitance of the circuit ($\Ctot = \Cm + \Cl + \Cs$). In addition to the motional capacitance, the total circuit capacitance consists of the intrinsic self-capacitance of the inductor coil ($C_l$) and any additional stray capacitance of the circuit ($C_s$). This ratio - and hence the electromechanical coupling - is maximized for small $\Cs+\Cl$ and large $\Cm$. We achieve a relatively small value of coil capacitance by using a tightly wound rectangular spiral inductor~\cite{Mohan1999,Fink2015a} with wire width of $550$~nm and wire-to-wire pitch of $1$~$\mu$m.  A simulation of the entire circuit layout, including nearby ground-plane, coupling wire, and connecting wires between the coil and the motional capacitor yields an additional stray capacitance of $\Cs = 1.13$~fF.  For a coil of $34$ turns, with an estimated inductance of $L = 46.3$~nH and capacitance $\Cl = 3.05$~fF, connected in parallel to a motional capacitance of $\Cm = 2.76$~fF corresponding to a vacuum gap of $d = 60$~nm, the microwave resonance frequency of the coupled circuit is estimated to be $\omegacE/2\pi = 8.88$~GHz. Using these circuit parameters in conjunction with a perturbative calculation~\cite{Eichenfield2009,Pitanti2015} of $(1/\Cm)\partial \Cm/\partial u$ based upon FEM simulations of the differential mechanical mode and the electric field distribution in the vacuum gap capacitor, yields a calculated vacuum electromechanical coupling strength of $\gzeroE/2\pi = 29.3$~Hz.  The trend of both $\Cm$ and $\gzeroE$ with gap size $d$ are shown in Fig.~\ref{fig:design}(c).

The devices studied in this work are fabricated from $1$~cm $\times$ $1$~cm chips diced from a high-resistivity silicon-on-insulator (SOI) wafer manufactured by SOITEC using the Smart Cut process~\cite{SOITEC}. The SOI wafer consists of a $300$~nm thick silicon device layer with (100) surface orientation and $p$-type (Boron) doping with a specified resistivity of $500$~$\Omega$-cm.  Underneath the device layer is a $3$~$\mu$m buried silicon dioxide ($\SiO$) BOX layer. The device and BOX layers sit atop a silicon (Si) handle wafer of thickness $675$~$\mu$m and a specified resistivity of $750$~$\Omega$-cm.  Both the Si device layer and handle wafer are grown using the Czochralski crystal growth method.  


Fabrication of the coupled coil resonator and H-slot resonator can be broken down into the following six steps.  In step (1), we pattern the H-slot resonator using electron beam (e-beam) lithography in ZEP-520A resist, and etch this pattern into the Si device layer using an inductively coupled plasma reactive ion etch (ICP-RIE). After the ICP-RIE etch, we clean the chips with a $4$~min piranha bath and a $12$~sec buffered hydrofluoric acid (BHF) dip. In step (2), we pattern the capacitor electrodes and ground plane region using ZEP-520A resist and use electron beam evaporation to deposit $60$~nm of Al on the chip.  In step (3), we define a protective scaffold formed out of LOR 5B e-beam resist to create the crossover regions of the spiral inductor coil.  In step (4), we pattern the inductor coil wiring in a double stack of PMMA 495 and PMMA 950 resists and deposit $120$~nm of Al using electron beam evaporation.  In step (5), we define a metal contact region that connects the wiring between the capacitor electrodes and the inductive coil, then perform a $5$~min ion mill before evaporating $140$~nm of Al. After all metal layer depositions, we perform a lift-off process for $1$~h in N-Methyl-2-pyrrolidone at 150$^o$C. 

In a final step (6), we release the structure by using an anhydrous vapor hydrofluoric (HF) acid etch using the SPTS uEtch system.  This etch is used to selectively remove the underlying BOX layer without attacking the Al metal or Si device layers.  Not only is the removal of the $\SiO$ BOX layer desirable from the standpoint of allowing the mechanical structure to move, but we have found that the presence of the underlying BOX layer results in a significant amount of electrical loss in the microwave resonator.  Measurements of both co-planar waveguide and lumped element microwave resonators have shown that the microwave $Q$-factor is substantially degraded (resonances difficult to detect; $\Qr \lesssim 100$) with the BOX layer present.  Stripping off the Si device layer and forming microwave resonators directly on the BOX layer marginally improves the microwave $Q$-factor ($\Qr \approx 300$), whereas stripping off both the device layer and BOX layer realize microwave resonators with $\Qr \approx 4 \times 10^{4}$ when fabricated directly on the Si handle wafer.  The release of the structure is facilitated by patterning an array of small ($100$~nm diameter) holes into the Si device layer during step (1).  The array of release holes are on a $2$~$\mu$m pitch and cover the region containing the coil and H-slot resonator. A timed etch of $75$~min is used to remove $\approx$~6~$\mu$m of $\SiO$, resulting in complete removal of the BOX layer underneath the microwave circuit.  A scanning electron microscope (SEM) image of the fully released structure is shown in Fig.~\ref{fig:fab}(a).  Zoom-in images of the H-slot mechanical resonator and undercut inductor coil are shown in Figs.~\ref{fig:fab}(b) and (c), respectively.

\begin{figure*}[btp]
\begin{center}
\includegraphics[width=\textwidth]{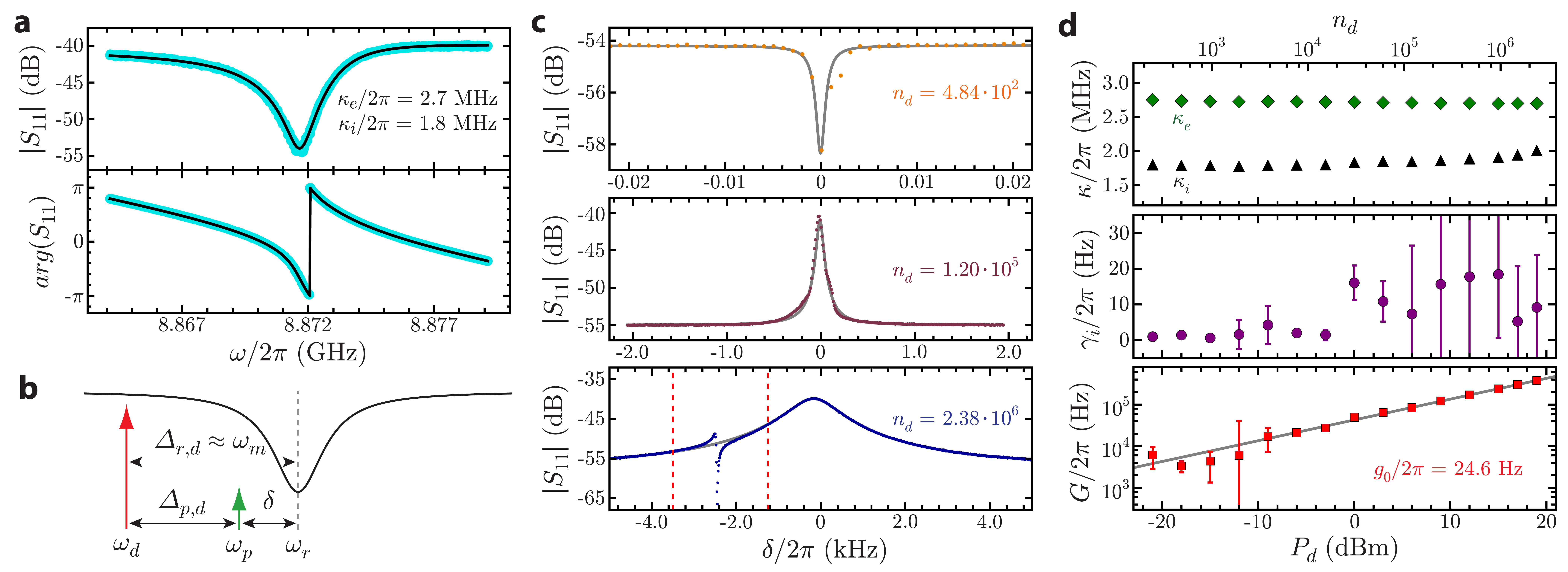}
\caption{\textbf{a}, Phase and amplitude response of the microwave resonator at fridge temperature $\Tf \approx 11$~mK and on-resonance cavity photon number of $\nprobe = 3.3$. The intrinsic loss rate $\kappaEi$ and external coupling rate $\kappaEe$ are extracted by fitting the curves with a modified Lorentzian cavity model to take into account the asymmetry in the background frequency response.  \textbf{b}, Schematic showing two-tone EIT measurement procedure. A strong drive tone at frequency $\omegad$ is placed on the red sideband of the microwave cavity and the cavity response is swept by a weak VNA probe at $\omega_p$. \textbf{c}, Plot of the measured EIT spectra at a series of drive intra-cavity photon numbers for a fridge temperature of $\Tf \approx 11$~mK.  From top to bottom: $\ndrive = 484$ (orange curve), $1.20 \times 10^5$ (maroon curve), and $2.38\times 10^6$ (blue curve).  Note for the blue curve at $\ndrive=2.38\times 10^6$, a weakly coupled auxiliary mechanical mode can be observed.  The frequency range between the vertical red dashed lines, surrounding the auxiliary mechanical resonance, were omitted for fitting purposes.  \textbf{d}, Plot of the fit values from the measured EIT spectra using Eq.~(\ref{eq:EIT}) for cavity coupling rates (top), intrinsic damping (middle), and parametrically enhanced coupling rate (bottom).  Error bars correspond to a $95\%$ confidence interval in the estimated fit parameter.}  \label{fig:char}
\end{center} 
\end{figure*}

Electromechanical measurements of the fabricated coil resonator are performed in a dilution refrigerator down to a temperature of $\Tf \approx 11$~mK.  Microwave signals are launched onto the SOI chip using a $50$~$\Omega$ co-planar waveguide.  The co-planar waveguide is terminated by extending the center conductor with a $2$~$\mu$m wide wire and then shorting it into ground.  The wire is passed within $9$~$\mu$m of the side of the inductor coil [see Fig.~\ref{fig:fab}(a)], thus providing large inductive coupling to the microwave resonator.  A region extending roughly $10$~$\mu$m from the edge of the surrounding ground plane of the co-planar waveguide an inductive coupling wire is also undercut and the BOX layer fully removed.  Read-out of the reflected microwave signal is performed using the measurement scheme shown in Fig.~\ref{fig:design}(a). The input line is thermalized at each stage of the fridge with a series of attenuators to eliminate Johnson thermal noise from the room temperature environment, yielding a calibrated input line attenuation of $\mathcal{A} = -73.9$~dB.  The reflected signal is redirected using a pair of circulators at the mixing chamber stage of the fridge and then sent into an amplifier chain consisting of a HEMT amplifier at the $4$~K fridge stage and a low-noise amplifier at room temperature.  The total amplifier gain is $52$~dB with an equivalent added microwave noise photon number of $\nadd \approx 30$.  

Figure~\ref{fig:char}(a) shows the measured magnitude and phase of the reflected microwave signal versus frequency from a vector network analyzer (VNA) used to probe the electrical properties of the device.  The microwave resonance frequency is measured to be $\omegacE/2\pi = 8.872$~GHz, in close correspondence to the resonance frequency based upon the simulated values of the coil inductance and the motional and stray capacitance of the circuit.  At the lowest base temperature of our fridge, $\Tf \approx 11$~mK, we measure an intrinsic microwave cavity loss rate of $\kappaEi/2\pi = 1.8$~MHz at an intra-cavity photon number on resonance of $\nprobe = 3.3$, corresponding to an internal quality factor of $\Qri = 4890$. The external coupling rate to the resonator is measured to be $\kappaEe = 2.7$~MHz, putting the device well into the overcoupled regime. We note that for similar coil resonators (without an H-slot resonator and coil cross-overs) which were coupled more weakly using a transmission as opposed to reflection geometry, we have observed internal quality factors as high as $\Qri \approx 2\times 10^4$, close to the measured $Q$-values for resonators fabricated directly on the Si handle wafer.  Further investigation is needed to determine the source of the additional microwave loss in the electro-mechanical devices studied here.


To characterize the mechanical properties of the H-slot resonator, and to determine the strength of its radiation pressure coupling to the microwave coil resonator, we perform two-tone pump and probe measurements as illustrated in Fig.~\ref{fig:char}(b). Here, a strong drive tone of power $\Pdrive$ is applied at frequency $\omegad$ on the red motional sideband of the microwave cavity resonance while a weak probe tone is scanned across the cavity resonance.  Interference between the anti-Stokes sideband of the drive tone and the weak probe tone results in a form of mechanically-mediated electromagnetically-induced transparency (EIT)~\cite{Weis2010,Chang2011,Teufel2011,Safavi-Naeini2011}, which for pump detuning near two-photon resonance ($\Deltard \equiv \omegacE - \omegad \approx \omegam$) yields a reflection spectrum given by, 

\begin{equation}
S_{11}(\delta) = 1- \frac{\kappaEe}{\kappaE/2 + i\delta + \frac{2\GOM^2}{\gammai + i2(\delta-(\omegam-\Deltard))}},    
\label{eq:EIT}
\end{equation}

\noindent where $\delta \equiv \omegap - \omegacE$ is the detuning of the probe frequency ($\omegap$) from the cavity resonance ($\omegacE$), and $\kappaE=\kappaEi + \kappaEe$ is the total loaded cavity damping rate.  Here we have made approximations assuming the system is sideband resolved ($\omegam/\kappaE \gg 1$) and that the probe signal is weak enough so as to not saturate the drive tone~\cite{Safavi-Naeini2011}.  

A subset of the measured spectra over a range of drive powers are shown in Fig.~\ref{fig:char}(c) around the EIT transparency window.  The drive detuning at two-photon resonance corresponds to the mechanical resonance frequency, and is found to be $\omegam/2\pi = 9.685$~MHz, very close to the numerically simulated resonance frequency of the in-plane differential mode of the H-slot resonator.  The cooperativity associated with the coupling of the microwave cavity field to the mechanical resonator is given by $C \equiv 4\GOM^2/\kappaE\gammai$, where $\gammaOME \equiv 4\GOM^2/\kappaE = 4\ndrive\gzeroE^2/\kappa$ is the back-action-induced damping of the mechanical resonator by the microwave drive tone.  At low drive powers corresponding to $C < 1$, we observe a narrow dip at the center of the broad microwave cavity resonance.  As the drive power is increased and $C > 1$ the dip becomes a peak in the reflected signal and the bandwidth of the transparency window increases with pump power.  At the highest powers we observe a substantially broadened transparency window, where we observe the presence of a second, spurious mechanical resonance about $2.4$~kHz below that of the strongly coupled resonance.  We attribute this spurious resonance to weak hybridization of the extended membrane modes of the undercut SOI with the localized in-plane differential mode of the H-slot resonator.  

Ignoring the spurious mechanical mode, we fit the measured EIT spectra using Eq.~(\ref{eq:EIT}) and extract the microwave cavity parameters ($\kappaEi$, $\kappaEe$, $\omegacE$), the intrinsic mechanical damping ($\gammai$), the mechanical resonance frequency ($\omegam$), and the parametrically-enhanced electromechanical coupling rate ($\GOM = \sqrt{\ndrive}\gzeroE$).  Figure~\ref{fig:char}(d) plots each of these fit parameters versus drive power and intra-cavity drive photon number ($\ndrive$).  $\kappaEi$ is found to weakly rise with $\ndrive$, most likely due to heating of the SOI membrane resulting from absorption of the microwave pump.  For intra-cavity photon number $\ndrive \gtrsim 5 \times 10^6$, we no longer observe a microwave resonance, suggesting that absorption of the microwave pump causes the superconducting circuit to go normal.  Fitting the measured curve of $\GOM$ versus pump photon number yields an estimate for the vacuum electromechanical coupling rate of $\gzeroE/2\pi = 24.6$~Hz, in good correspondence with the simulated value of $29.3$~Hz.  The slight discrepancy is likely attributable to an under-estimation of the true capacitor gap size due to overhang of the Al electrode into the gap.  For a $70$~nm dielectric gap, consistent with an additional $10$~nm overhang of Al estimated from cross-sectional images of similar devices, the simulated vacuum coupling rate drops to $\gzeroE/2\pi = 24.4$~Hz   

\begin{figure}[t]
\begin{center}
\includegraphics[width=\columnwidth]{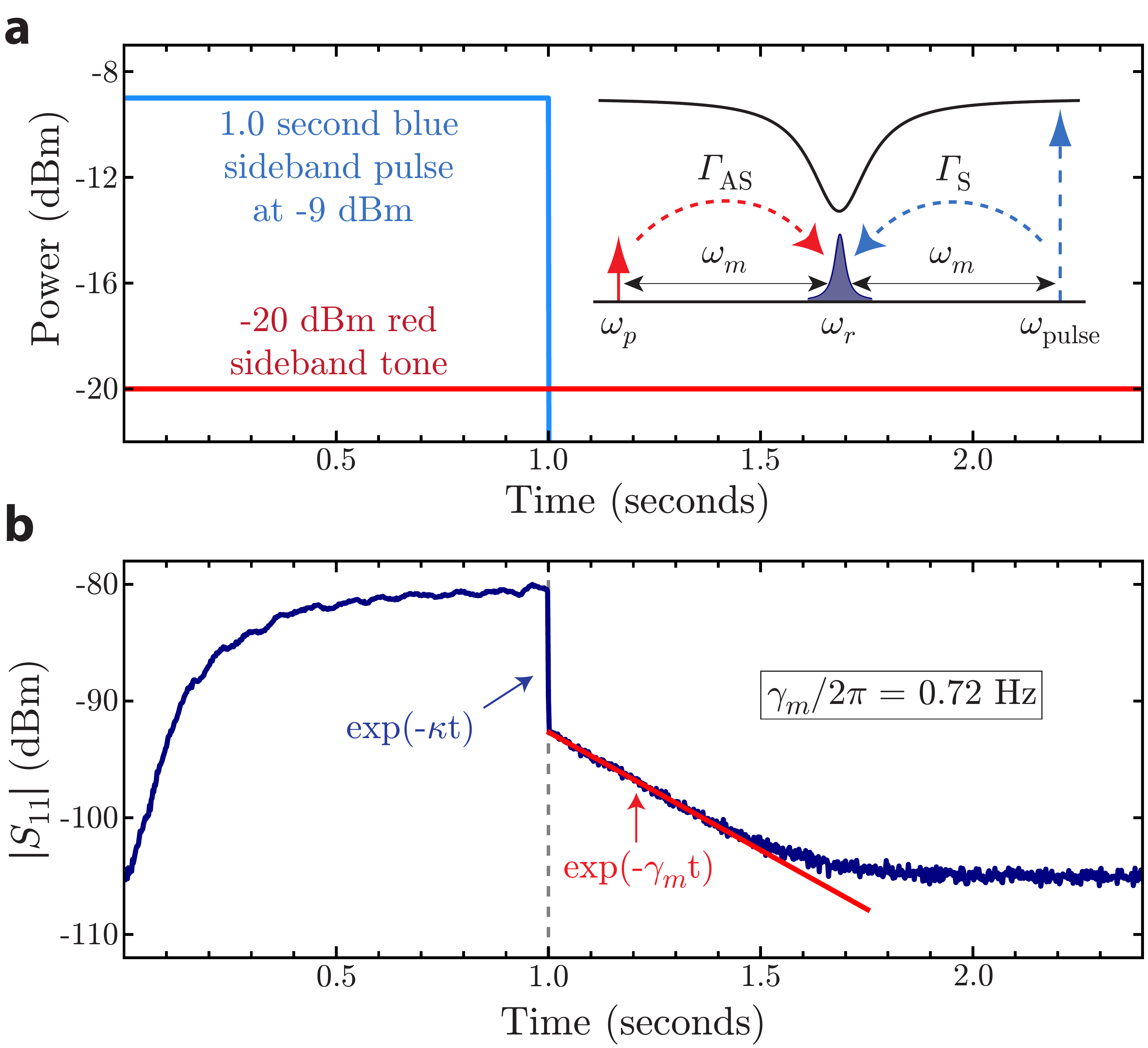}
\caption{\textbf{a}, Schematic showing the time-domain mechanical ring-down protocol, wherein a strong blue pulse at $\omegapulse = \omegacE+\omegam$ populates the mechanics and a weak probe tone at $\omegap = \omegacE-\omegam$ is used to monitor the energy in mechanical resonator. Inset: schematic showing the frequency and scattering of the applied tones used to ring-up and monitor the mechanical resonator. Here, $\GammaS \approx (4\npulse\gzeroE^2/\kappa)(\nm+1)$ [$\GammaAS \approx (4\nprobe\gzeroE^2/\kappa)\nm$] is the Stokes [anti-Stokes] scattering rate of the pulse [probe] tone, where $\npulse$ [$\nprobe$] is the intra-cavity pulse [probe] tone photon number. \textbf{b}, Time-domain mechanical ring-down measurement at $\Tf \approx 11$~mK. A steep decay resulting from the leakage of photons from the cavity is followed by a slow decay dominated the intrinsic mechanical damping of the resonator is observed.} \label{fig:ringdown}
\end{center} 
\end{figure}

At low drive powers, the EIT transparency window bandwidth is dominated by the intrinsic mechanical linewidth, $\gammai$. However, we observe a frequency jitter of the mechanical resonance frequency.  The range of the frequency jitter is of the order of several Hz on the second timescale, and saturates at approximately $20$~Hz for timescales on the order of tens of minutes. The source of the mechanical frequency jitter is unknown, but may be related to the excess heating we observe at the lowest fridge temperatures as described below.  The resolution bandwidth of the VNA is also limited to $1$~Hz, and combined with the frequency jitter leads to significant distortion and blurring of the measured EIT spectrum at low drive power as can be seen in the top plot of Fig.~\ref{fig:char}(c). Estimates of $\gammai$ and the corresponding cooperativity $C$ are thus unreliable from the EIT spectra alone. 


To directly measure the intrinsic damping rate $\gammai$ we perform a mechanical ring-down measurement as shown in Fig.~\ref{fig:ringdown}. Here, a strong blue-detuned pulse tone at frequency $\omegapulse = \omegacE + \omegam$ is applied for $1$~s to amplify the thermal mechanical motion of the mechanical resonator through dynamic back-action~\cite{Kippenberg2005,Grudinin2010,Cohen2015}.  A weak red-detuned probe tone at $\omegap = \omegacE - \omegam$ is applied to the microwave cavity in order to read-out the phonon occupancy of the resonator after the blue detuned pulse is turned off.  A spectrum analyzer with resolution bandwidth set to RBW$=1$~kHz is used to measure the motionally scattered photons near the cavity resonance from the pulse and probe tones, providing a time domain signal proportional to the mechanical resonator phonon occupancy as shown in Fig.~\ref{fig:ringdown}(b).  

Fitting the decay of the spectrum analyzer signal after the pulse tone is turned off, and after the initial rapid decay of pulse photons from the cavity, yields a mechanical damping rate of $\gammam/2\pi = 0.72$~Hz.  Note that as the probe tone power of $-20$~dBm corresponds to an intra-cavity photon number of only $\nprobe \approx 300$, the dynamic back-action damping of the probe is small but non-negligible at $\gammaOME/2\pi \approx 0.16$~Hz.  The corresponding intrinsic mechanical damping rate is thus approximately, $\gammai/2\pi \approx 0.56$~Hz, corresponding to a mechanical quality factor of $\Qm = 1.7\times 10^7$.  


\begin{figure}[htp!]
\begin{center}
\includegraphics[width=\columnwidth]{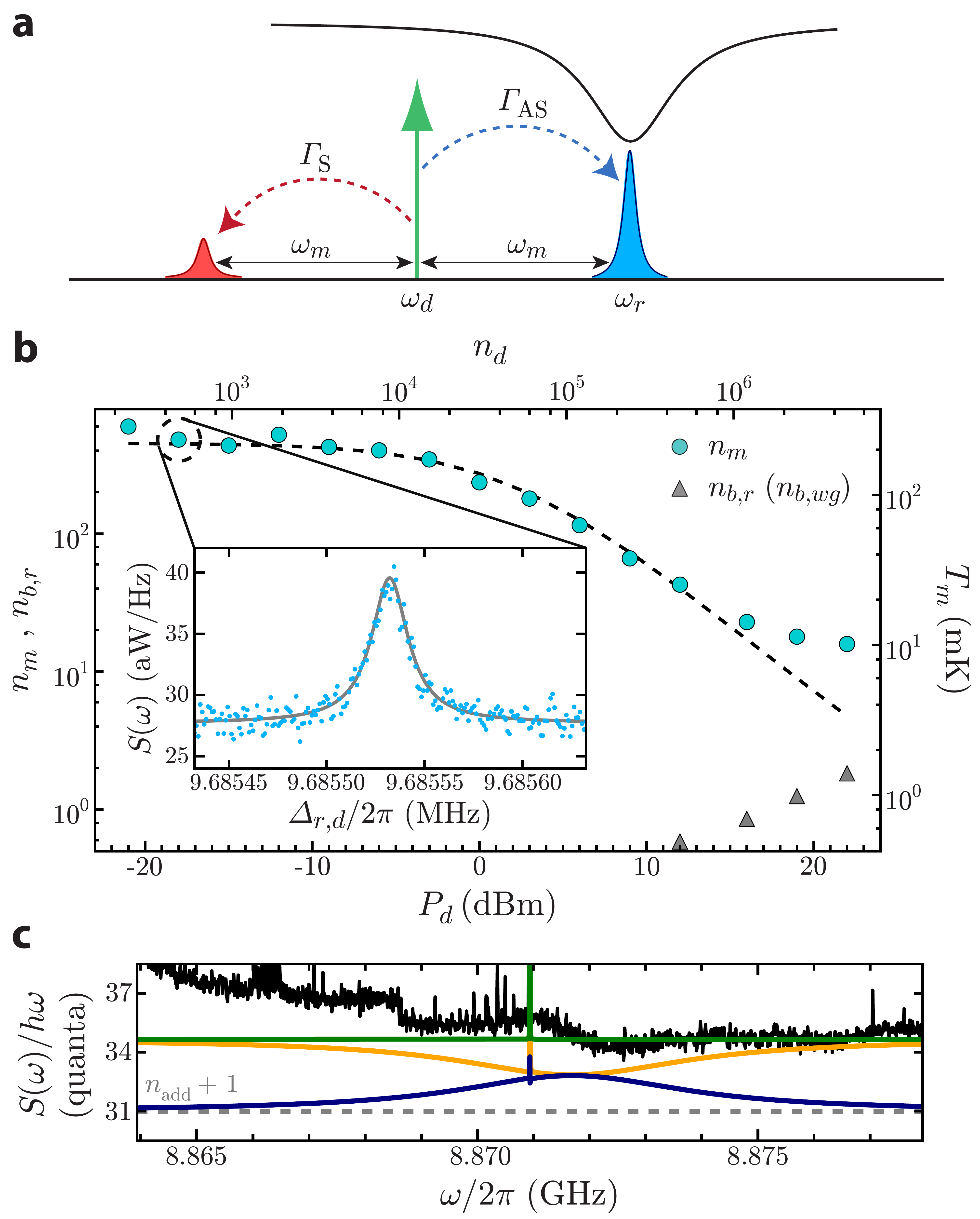}
\caption{\textbf{a}, Schematic showing thermometry measurement scheme, wherein a red detuned drive tone is used to simultaneously cool and measure the average energy in the mechanical resonator through anti-Stokes scattering proportional to phonon occupancy of the resonator. $\GammaAS \approx (4\ndrive\gzeroE^2/\kappa)\nm$ is the cavity-enhanced anti-Stokes scattering rate of the drive tone proportional to $\nm$.  $\GammaS \approx (4\ndrive\gzeroE^2/\kappa)(\kappa/4\omegam)^2(\nm+1)$ is the cavity-supressed Stokes scattering rate of the drive tone proportional to $\nm+1$.  \textbf{b}, Cooling curve obtained by fitting the measured microwave noise spectrum using a model which includes noise squashing and heating effects due to thermal noise in the microwave cavity and the input coupler. Spectra are taken at a fridge temperature of $\Tf = 211$~mK. Blue circles correspond to the inferred average mechanical mode occupancy ($\nm$) from fits to the measured noise (see inset).  Grey triangles are the fit input waveguide ($\nwb$) and cavity ($\nrb$) thermal noise occupancies from the measured noise background level.  The dashed line indicates the predicted occupancy as given by $\nfm/(1+C)$, where $C$ is determined from the EIT fit values for the vacuum coupling rate ($\gzeroE/2\pi=25.1$~Hz) and the intrinsic damping rate ($\gammai/2\pi = 25.7$~Hz) taken at a fridge temperature of $\Tf \approx 211$~mK. \textbf{c}, Plot of the measured noise spectral density (black curve) and modeled noise background (green curve) at $\Pdrive = 22$~dBm. The orange curve corresponds to the expected spectral noise density due to the waveguide bath ($\nwb$) alone while the navy curve shows the expected contribution from the resonator bath ($\nrb$).  The global offset of $\nadd + 1$ shown as a grey dashed line.} \label{fig:cooling}
\end{center} 
\end{figure}

The red-sideband pump configuration ($\Deltard = \omegam$) used to measure the EIT spectra also cools the mechanical resonator.  Using a spectrum analyzer to measure the anti-Stokes scattered drive photons near the microwave cavity resonance, as shown schematically in Fig.~\ref{fig:cooling}(a), and calibrating the measured noise spectrum allows one to infer the average (noise/thermal) phonon occupancy ($\nm$) of the mechanical resonator as a function of the drive power.  This sort of mechanical mode thermometry~\cite{Teufel2011,Fink2015a} at the lowest fridge temperature of $\Tf \approx 11$~mK shows large fluctuations in the inferred mechanical mode temperature ($\Tm = 20-200$~mK), as a function of both time and drive power.  The source of this anomalous mechanical mode heating is not well understood at this point, but may be related to coupling between the driven microwave resonator and two-level systems (TLS)~\cite{GaoPhD} present in the native oxide on the surface of the Si device~\cite{Gao2008,Wisbey2010,Bruno2015}.  TLS can not only absorb energy from the microwave drive, but also may hybridize with the microwave cavity and influence the transduction of mechanical motion yielding artificially high or low inferred mechanical mode occupancy~\cite{Fink2015a}.   

To better characterize the back-action cooling of the mechanics, we performed a cooling sweep at a fridge temperature of $\Tf \approx 211$~mK, where the anomalous heating effects seen at $\Tf \approx 11$~mK are less significant in comparison to the thermal bath of the fridge.  A plot of the inferred mechanical mode occupancy ($\nm$) and corresponding mode temperature ($\Tm$) versus the drive power applied on the red-sideband ($\Deltard \approx \omegam$) is shown in Fig.~\ref{fig:cooling}(b).  At low drive powers where $C \ll 1$ and back-action cooling is expected to be negligible, we find that the mechanics thermalizes to an occupancy very close to the mechanical thermal occupancy at the fridge temperature, $\nfm = 453$.  For comparison a plot of the ideal cooling curve, $\nm = \nfm/(1+C)$, due to radiation pressure back-action is shown as a dashed curve in Fig.~\ref{fig:cooling}(b).  Here we use an intrinsic damping rate ($\gammai/2\pi = 25.7$~Hz) and vacuum coupling rate ($\gzeroE/2\pi = 25.1$~Hz) inferred from EIT measurements at $\Tf \approx 211$~mK.  The measured mechanical mode occupancy is in good agreement with theory, except at powers $\Pdrive > 10$~dBm where we again observe anomalous heating effects. Nonetheless, we are able to perform well over a decade of cooling and reach occupancies as low as $\nm \approx 16$. 

In addition to the measured Lorentzian noise peak of the mechanical resonance, we also observe broadband noise which increases with the drive power.  Figure~\ref{fig:cooling}(c) shows a plot of the measured broadband noise (back curve) at the highest intra-cavity drive photon number of $\ndrive = 4.75 \times 10^{6}$.  This broadband noise does not seem to be phase noise of our microwave source as addition of a narrowband input filter had no effect on the measured noise spectrum.  Assuming that the noise is associated with an elevated electrical noise temperature of the device, we include both an input waveguide thermal noise occupancy ($\nwb$) and a cavity thermal noise occupancy ($\nrb$) to our model. Taking the waveguide and cavity to be at the same noise temperature (i.e., $\nwb = \nrb$) yields a flat reflection noise spectrum as shown in Fig.~\ref{fig:cooling}(c).  Fitting the noise background at each drive power yields an estimate for the cavity and waveguide noise photon numbers, which are shown versus drive power as grey triangles in Fig.~\ref{fig:cooling}(b).  The inferred effective noise temperature of the microwave cavity at the highest drive power is $\Tr \approx 1$~K, close to the critical temperature for Al~\cite{Cochran1958} and consistent with the circuit going normal at higher drive powers.   

Enhancement in the back-action cooling and electro-mechanical cooperativity of the current devices can be realized most straightforwardly through reduction in the microwave resonator loss.  Significant reduction in the microwave loss and heating effects should be attainable through the use of higher resistivity Si~\cite{Wisbey2010,Bruno2015}.  The vacuum electro-mechanical coupling rate may also be increased to $\gzeroE/2\pi \approx 100$~Hz through optimization of the circuit layout to reduce stray capacitance and reduction of the capacitor gap to previously reported values of $d \approx 30$~nm~\cite{Pitanti2015}.  As the Si dielectric loss is reduced, and if one hopes to create highly coherent qubits on SOI~\cite{Weber2011,Patel2013}, TLS in the Si surfaces will need to be mitigated~\cite{Gao2008,Wisbey2010,Bruno2015}.  In this regard, anhydrous vapor HF etching may be useful as it is compatible with numerous electron beam and photosensitive resists, thus allowing cleaning and hydrogen passivation of the Si surface right before and after critical fabrication steps of superconducting resonators and Josephson-Junction qubits~\cite{Oliver2013}.  

Looking forward, SOI represents a unique platform for integrating microwave, mechanical, and optical circuits. This is particularly interesting in the context of recent proposals and experimental efforts to utilize mechanical elements as quantum converters between microwave and optical light~\cite{Safavi-Naeini2011,Wang2012b,Bochmann2013,Andrews2014,Bagci2014,Cernotik2015}.  SOI is the substrate of choice for many nano- and micro-photonics applications due to the advanced fabrication techniques developed in Si and its excellent waveguiding and optical properties in the $1550$~nm telecom wavelength band~\cite{Bogaerts2005}.  SOI has also been successfully utilized in the development of thin-film optomechanical crystals, in which photonic and phononic bandgap effects are utilized to co-localize and strongly interact near-IR light with GHz-frequency phonons~\cite{Eichenfield2009,Meenehan2015}.  Such a device architecture would allow for a fully integrated, scalable platform for on-chip quantum information processing and long-distance quantum networking~\cite{Kimble2008}.  



\begin{acknowledgments}
The authors would like to thank Dan Vestyck at SPTS for performing trial HF vapor etch fabrication runs, and Barry Baker for his good humor and tireless effort to get the HF vapor etcher set up at Caltech. The authors also thank Alessandro Pitanti and Richard Norte for their contributions to initial fabrication development in SOI. This work was supported by the AFOSR through the ``Wiring Quantum Networks with Mechanical Transducers'' MURI program, the Institute for Quantum Information and Matter, an NSF Physics Frontiers Center with support of the Gordon and Betty Moore Foundation, and the Kavli Nanoscience Institute at Caltech. P.B.D. acknowledges support from a Barry Goldwater Scholarship.
\end{acknowledgments}

%

\end{document}